\documentclass[a4paper,12pt]{article}

\usepackage{ifpdf}

\newif\ifpdf
\ifx\pdfoutput\undefined
  \pdffalse
\else
  \pdfoutput=1
  \pdftrue
\fi

\RequirePackage{xspace} %
\RequirePackage{subfigure} %
\RequirePackage[centertags]{amsmath} %
\RequirePackage{amssymb}
\RequirePackage{wrapfig} %
\RequirePackage{calc} %
\RequirePackage{ifthen}
\RequirePackage{tabularx} %
\RequirePackage{flafter} %
\RequirePackage{fancyhdr} %

\ifpdf
  \RequirePackage[pdftex]{color}%
  \RequirePackage{colortbl}%
  \RequirePackage{array}%
  \RequirePackage[pdftex]{graphicx}

  \RequirePackage[ pdftex, plainpages = false, pdfpagelabels,
                 pdfpagelayout = useoutlines,
                 bookmarks,
                 breaklinks = true,
                 linktocpage,
                 pagebackref,                      % to include page numbers in bibliography
                 colorlinks = true,
                 linkcolor = blue,
                 urlcolor  = blue,
                 citecolor = blue,
                 anchorcolor = blue,
                 hyperindex = true,
                 hyperfigures
                 ]{hyperref}

\else
  \RequirePackage{color}
  \RequirePackage{colortbl}
   \RequirePackage{array}
  \RequirePackage[dvips]{graphicx}
  \RequirePackage{hyperref}
  \usepackage{rotating}
\fi

\usepackage{makeidx} % enable indexing
\usepackage{setspace} %enable setting spacing, %e.g. singlespace, onehalfspace, and doublespace
\usepackage{rotating} %enable sidewaysfigure
\usepackage{ecltree}
\usepackage{epic}
\usepackage{supertabular}  % this will allow the table not to break
\usepackage{color}
\usepackage{exscale}
\usepackage{fontenc}
\usepackage{ifthen}
\usepackage{latexsym}
\usepackage{makeidx}
\usepackage{syntonly}
\usepackage{inputenc}
\usepackage{graphicx}
\usepackage{setspace}
\usepackage{caption2}
\usepackage[english]{babel}
\usepackage[square, comma,numbers,sort&compress]{natbib}
\usepackage{hypernat}
\usepackage{boxedminipage}
\usepackage{framed}
\usepackage{longtable}
\usepackage[all]{hypcap}    %included to make the hyperlink go to the top of figure
\usepackage{algorithm2e}
\usepackage{algorithmic}
\usepackage{lscape}
\usepackage{pdflscape}
\newcommand{\note}[1]{\marginpar[left]{\singlespace \tiny #1}}
\newcommand{\pois}{Poiseuille}
\newcommand{\Vis}    {\mu}            % Viscosity
\newcommand{\lVis}   {\mu_{0}}        % Low-shear viscosity
\newcommand{\sR}     {\dot{\gamma}}         % Shear rate
\newcommand{\sS}     {\tau}           % Shear stress
\newcommand{\ysS}    {\sS_{o}}        % Yield stress
\newcommand{\sTenC}  {\tau}           % Stress tensor component
\newcommand{\te}[1]{\mbox{\boldmath{$#1$}}}                    %tensor: boldface
\newcommand{\ucd}[1]{\overset{\bigtriangledown}{\mbox{$#1$}}}
\newcommand{\Tim}    {\lambda}        % Time constant
\newcommand{\rxTim}  {\Tim_{1}}       % Relaxation time
\newcommand{\rdTim}  {\Tim_{2}}       % Retardation time
\newcommand{\sTen}   {\te \sTenC}     % Stress tensor (replace "\tau" with "sS" ?)
\newcommand{\rsTenC}  {\dot{\gamma}}        % Rate-of-strain tensor component
\newcommand{\rsTen}  {\te \rsTenC}    % Rate-of-strain tensor (replace "\gamma" with "sR" ?)

\renewcommand{\sectionmark}[1]%
      {\markright{\thesection\ #1}} %stops it capitalizing. #1 has value of section name

\renewcommand{\note}[1]{}

\onehalfspace %\doublespace

\author{Taha Sochi\footnote{University College London, Department of Physics \& Astronomy, Gower Street, London, WC1E 6BT.
Email: t.sochi@ucl.ac.uk.} \vspace*{5.0cm}}

\date{}

\begin{document}
\begin{center}
{\Large Non-Newtonian Rheology in Blood Circulation}
\par\end{center}{\Large \par}

\begin{center}
Taha Sochi
\par\end{center}

\begin{center}
{\scriptsize University College London, Department of Physics \& Astronomy, Gower Street, London,
WC1E 6BT \\ Email: t.sochi@ucl.ac.uk.}
\par\end{center}

\begin{abstract}
\noindent Blood is a complex suspension that demonstrates several non-Newtonian rheological
characteristics such as deformation-rate dependency, viscoelasticity and yield stress. In this
paper we outline some issues related to the non-Newtonian effects in blood circulation system and
present modeling approaches based mostly on the past work in this field.

\vspace{0.3cm}

\noindent Keywords: hemorheology; hemodynamics; blood properties; biorheology; circulatory system;
fluid dynamics; non-Newtonian; shear thinning; yield stress; viscoelasticity; thixotropy.
\par\end{abstract}

\begin{center}

\par\end{center}

%XXXXXXXXXXXXXXXXXXXXXXXXXXXXXXXXXXXXXXXXXXXXXXXXXXXXXXXXXXXXXXXXX
\section{Introduction} \label{Introduction}

\begin{figure}[!h]
\centering{}
\includegraphics
[scale=0.4] {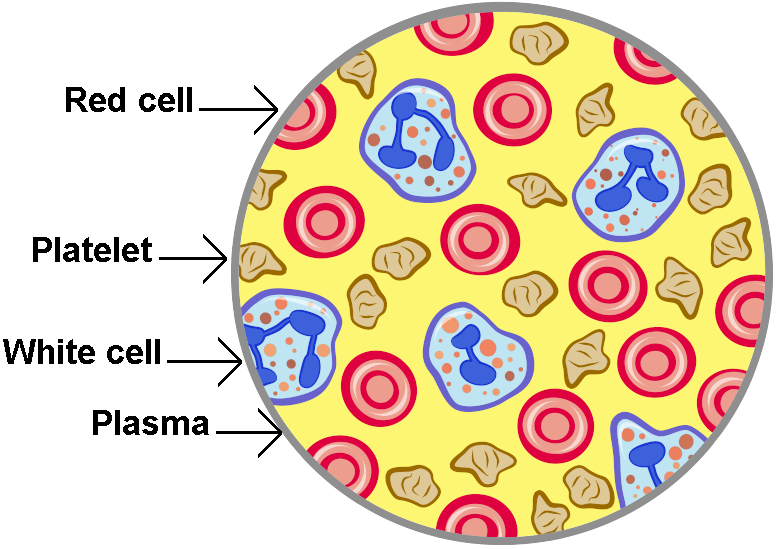} \caption{Simplified illustration of the main blood composition. The
number and size of cells shown in the figure are for demonstration purposes and not supposed to
reflect their proportion or relative size in the blood.} \label{BloodComponents}
\end{figure}

Blood is a heterogeneous multi-phase mixture of solid corpuscles (red blood cells, white blood
cells and platelets) suspended in a liquid plasma which is an aqueous solution of proteins, organic
molecules and minerals (refer to Figure \ref{BloodComponents}). The rheological characteristics of
blood are determined by the properties of these components and their interaction with each other as
well as with the surrounding structures. The blood rheology is also affected by the external
physical conditions such as temperature; however, in living organisms in general, and in large
mammals in particular, these conditions are regulated and hence they are subject to minor
variations that cannot affect the general properties significantly \cite{BodnarSP2011}. Other
physical properties, such as mass density, may also play a role in determining the blood overall
rheological conduct. The rheological properties of blood and blood vessels are affected by the body
intake of fluids, nutrients and medication \cite{GroglerLBB1997, VlastosTR2003, GapinskaGEJK2007,
TripetteLSGSe2010}, although in most cases the effect is not substantial except possibly over short
periods of time and normally does not have lasting consequences.

\begin{figure}[!h]
\centering{}
\includegraphics
[scale=1] {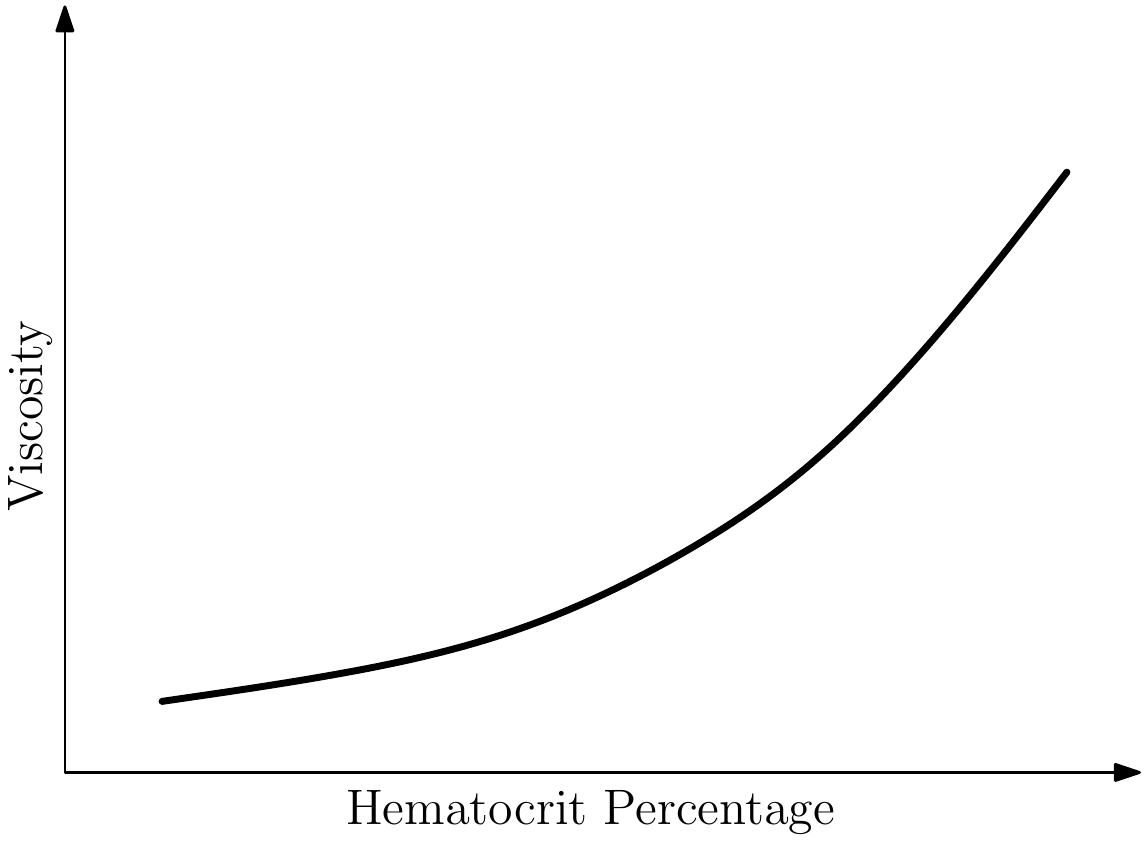} \caption{Dependency of human blood viscosity on hematocrit percentage
concentration on a linear-linear plot \cite{BaskurtM2003, LongUMD2005}.} \label{VisHemaPlot}
\end{figure}

The viscosity of blood is determined by several factors such as the viscosity of plasma, hematocrit
level (refer to Figures \ref{VisHemaPlot} and \ref{VisSRplot}), blood cell distribution, and the
mechanical properties of blood cells \cite{BaskurtM2003, FisherR2009, LeeXNLS2011}. The blood
viscosity is also affected by the applied deformation forces, extensional as well as shearing, and
the ambient physical conditions. While the plasma is essentially a Newtonian fluid, the blood as a
whole behaves as a non-Newtonian fluid showing all signs of non-Newtonian rheology which includes
deformation rate dependency, viscoelasticity, yield stress and thixotropy. Most non-Newtonian
effects originate from the red blood cells due to their high concentration and distinguished
mechanical properties such as elasticity and ability to aggregate forming three-dimensional
structures at low deformation rates \cite{Dintenfass1962, BodnarSP2011}.

Deep understanding of the blood rheology, which includes its non-Newtonian characteristics, is
important for both diagnosis and treatment. Looking into the existing biological literature and
comparing to the non-biological literature, such as earth science studies, it is obvious that the
non-Newtonian phenomena in the blood circulation has not been given sufficient attention in the
biological studies. One reason is the complexity of the biological systems which makes the
consideration of non-Newtonian effects in blood circulation more difficult to handle. Hence, to
simplify the flow models and their computational implementation, blood is generally assumed
Newtonian and these effects are ignored. The obvious difficulties in observation, experimentation
and measurement related to {\it in vivo} blood flow add another complicating factor. Moreover,
apart from some rare and extreme pathological states, the non-Newtonian effects in blood flow are
relatively mild as compared to the non-Newtonian effects exhibited by typical polymeric systems for
instance. This makes the approximation of blood as a Newtonian fluid an acceptable assumption and
not far from reality in a significant part of the circulatory system under normal conditions.

Many theoretical, numerical, experimental and clinical studies of non-Newtonian effects in blood
circulation have been conducted in the last few decades. However, there is no general approach in
tackling this problem in a systematic way based on a unified vision. Almost all the past studies
focus on individual problems and deal with the existing non-Newtonian phenomena within a limited
local context. The current study, which is basically a brief overview of this subject, is trying to
deal with the non-Newtonian blood rheology in general as applied to all levels of the circulation
system.

%XXXXXXXXXXXXXXXXXXXXXXXXXXXXXXXXXXXXXXXXXXXXXXXXXXXXXXXXXXXXXXXXX
\section{Non-Newtonian Characteristics}

As indicated already, blood is a complex non-Newtonian fluid showing various signs of non-Newtonian
rheology such as shear thinning \cite{GijsenAVJ1999}, yield stress \cite{MerrillCP1969,
Thurston1972, MorrisSB1987, FisherR2009, SochiYield2010, SochiFeature2010} and viscoelasticity
\cite{Merrill1969, Thurston1972, MorrisSB1987, KangBR2000, SochiVE2009, RevellinRBB2009,
BodnarSP2011}. The blood is also characterized by a distinctive thixotropic behavior
\cite{Dintenfass1962, HuangF1976} revealed by the appearance of hysteresis loops during shearing
cycles \cite{HuangF1976, Barnes1997, SochiArticle2010}. These non-Newtonian properties do not
affect the flow patterns inside the flow paths and the fluid transportation only but they also
affect the mechanical stress on the blood vessel walls and the surrounding tissues
\cite{ChenLW2006} especially in cases of irregular lumen geometry like stenosed arteries
\cite{LouY1993, LiuT2011, MisraM2012}. The mechanical stress on the vessel wall and tissue is not
only important for its direct mechanical impact, especially when sustained over long periods of
time, but it can also contribute to the commencement and advancement of long term lesions such as
forming sediments inside the vessel wall \cite{GolpayeghaniNM2008}. The non-Newtonian properties,
like viscoelasticity, have also an impact on other transport phenomena such as pulse wave
propagation in arteries \cite{KangBR2000}.

Non-Newtonian effects in general are dependent on the magnitude of deformation rates and hence they
can exist or be enhanced at certain flow regimes such as low shear rates \cite{FanJZLCD2009,
RevellinRBB2009}. The non-Newtonian effects are also influenced by the type of deformation, being
shear or elongation \cite{SochiFeature2010}. The impact of the non-Newtonian effects can be
amplified by a number of factors such as pathological blood rheology and flow in stenosed vessels
and stents \cite{Stoltz1985, IshikawaGOY1998, MandalML2012, HuangCS2013}. An interesting finding of
one study \cite{ShuklaPR1980} is that although flow resistance and wall shear stress increases as
the size of stenosis increases for a given rheological model, the non-Newtonian nature of blood
acts as a regulating factor to reduce the resistance and stress and hence contribute to the body
protection. In this context, shear thinning seems to have the most significant role in facilitating
blood flow through stenotic vessels.

\begin{figure}[!h]
\centering{}
\includegraphics
[scale=1] {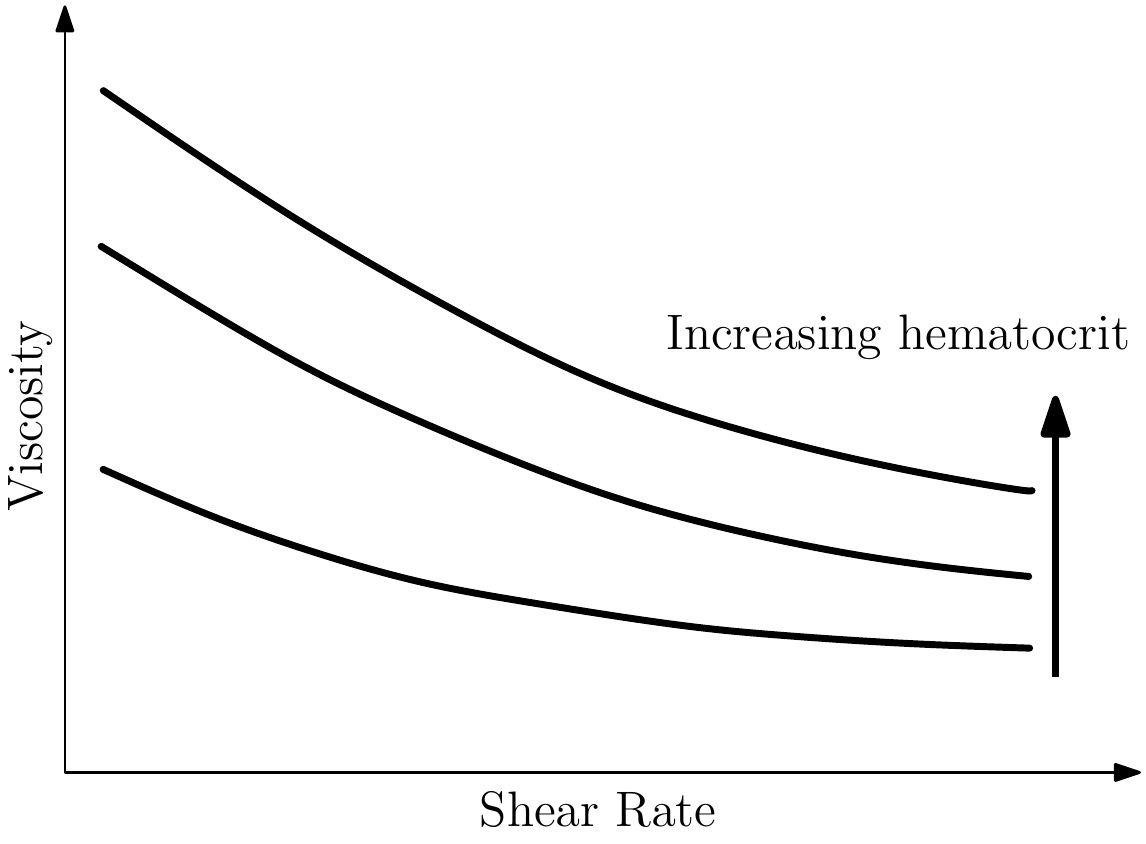} \caption{Human blood viscosity as a function of shear rate for a range of
hematocrit concentrations on a log-log plot \cite{DasJP2000, JungLPH2006, LeeXNLS2011}.}
\label{VisSRplot}
\end{figure}

Blood is a predominantly shear thinning fluid (refer to Figure \ref{VisSRplot}), especially under
steady flow conditions, and this property has the most important non-Newtonian impact
\cite{Mandal2005, ChenLW2006, RevellinRBB2009, FisherR2009}. Shear thinning is not a transient
characteristics; moreover it is demonstrated at most biological flow rates \cite{Mandal2005}
although it is more pronounced at low deformation regimes. Shear thinning rheology arises from
disaggregation of the red blood cells with increasing shear rate \cite{PerktoldKLH1999}. This same
reason is behind the observed thixotropic blood behavior as shearing forces steadily disrupt the
structured aggregation of blood cells with growing deformation time. The origin of other
non-Newtonian effects can also be traced back to the blood microstructure as will be discussed
next.

The viscoelastic nature of blood basically arises from its corpuscular microstructure. Viscoelastic
properties originate from the red blood cells, which are distinguished by their pronounced elastic
deformability \cite{HellBDB1989, PerktoldKLH1999} associated with the ability to aggregate forming
three-dimensional structures known as rouleaux \cite{Stoltz1985}. The aggregation is mostly
demonstrated at low shear rates and hence non-Newtonian behavior in general and viscoelasticity in
particular are more pronounced at these regimes of low deformation \cite{Pontrelli2000,
BaskurtM2003, BodnarSP2011}. The viscoelastic effects are magnified, if not activated, by the
pulsatile nature of blood flow. The viscoelastic effects in blood circulation should not be limited
to the viscoelastic properties of the blood itself but also to the viscoelastic (or elastic
depending on the adopted model) properties of the blood vessels and the porous tissue through which
the blood is transported \cite{CanicHRTGM2006}. This can be justified by the fact that all these
effects are manifested by the blood circulation and hence they participate, affect and affected by
the circulation process. An interaction between the viscoelastic behavior of blood with that of the
vessel wall and porous tissue is unavoidable consequence.

\begin{figure}[!h]
\centering{}
\includegraphics
[scale=1] {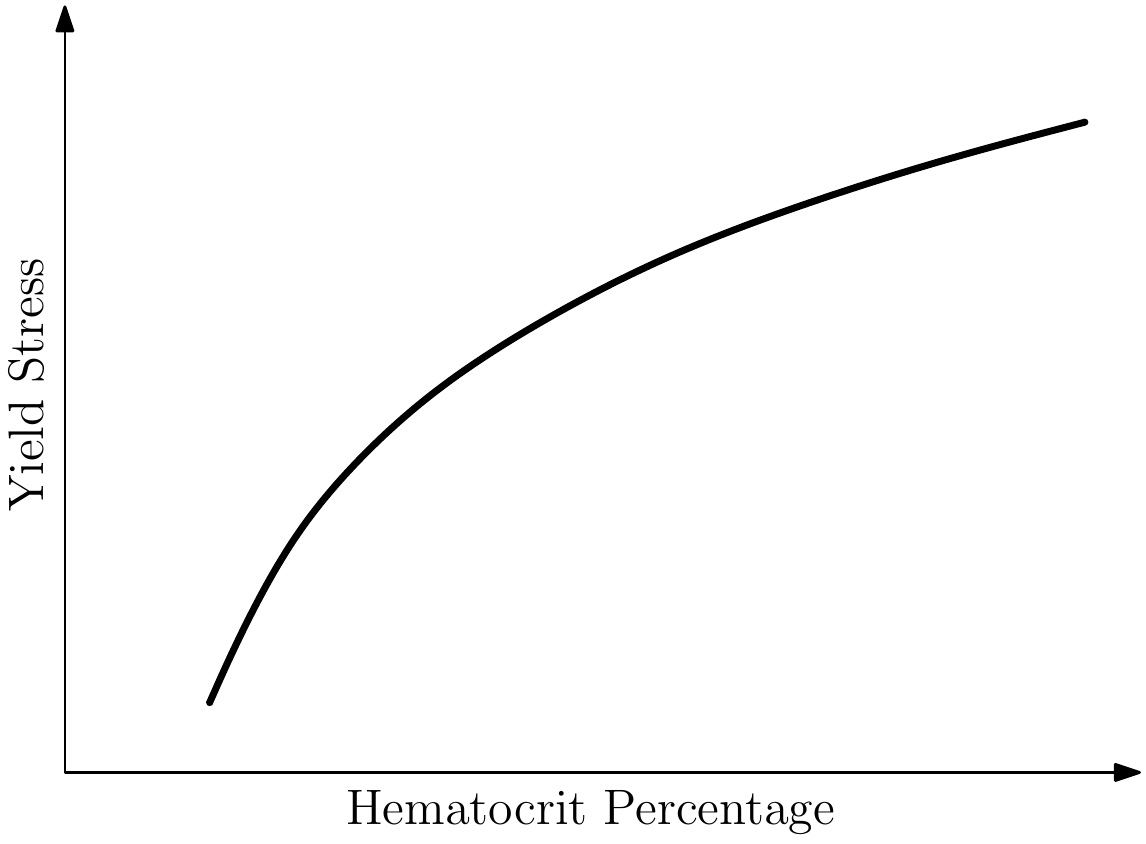} \caption{Dependence of yield stress of human blood on hematocrit level on
a log-linear plot according to some studies. Other types of correlation have also been obtained in
other studies \cite{MorrisSB1987, PicartPGC1998, LeeXNLS2011, FedosovPCGK2011}.} \label{YsHemaPlot}
\end{figure}

Blood also demonstrates yield stress although there is a controversy about this issue
\cite{LouY1993, BodnarSP2011}. Yield stress arises from the aggregation of red blood cells at low
shear rates to form the above-mentioned three-dimensional micro-structures (rouleaux) that resist
the flow \cite{ReplogleMM1967, FisherR2009, BodnarSP2011}. Studies have indicated that yield stress
is positively correlated to the concentration of fibrinogen protein in blood plasma and to the
hematocrit level \cite{Merrill1969, MorrisRSGSB1989, HellBDB1989, LouY1993}. An illustrative plot
of the dependence of yield stress on hematocrit level is shown in Figure \ref{YsHemaPlot}. Other
factors, such as the concentration of minerals, should also have a contribution. Many of the blood
rheological characteristics in general, and non-Newtonian in particular, are also controlled or
influenced by the fibrinogen level \cite{ReplogleMM1967}. The yield stress characteristic of blood
seems to vanish or become negligible when hematocrit level falls below a critical value
\cite{Merrill1969}. Yield stress contributes to the blood clotting following injuries and
subsequent healing, and may also contribute to the formation of blood clots (thrombosis) and vessel
blockage in some pathological cases such as strokes. The value of yield stress, as reported in a
number of clinical and experimental studies, seems to indicate that it is not significant and hence
it has no tangible effect on the flow profile (and hence flow rate) at the biological flow ranges
in large and medium size blood vessels \cite{LouY1993}. However, it should have more significant
impact in the minute capillaries and some porous structures where flow at very low shear rates
occurs. The magnitude of yield stress and its effect could be aggravated by certain diseased states
related to the rheology of blood, like polycythemia vera, or the structure of blood vessels such as
stenoses.

As a shear thinning fluid, blood is also characterized by a thixotropic behavior, which is
confirmed experimentally by a number of studies \cite{Dintenfass1962}, due to the intimate relation
between these two non-Newtonian properties \cite{SochiArticle2010}. This may also explain a
possible controversy about the thixotropic nature of blood \cite{DavenportR1981} as the
thixotropic-like behavior may be explained by other non-Newtonian characteristics of blood. Despite
the fact that thixotropy is a transient property, due to the pulsative nature of the blood flow the
thixotropic effects may have long term impact on the blood circulation. This equally applies to the
time-dependent effects of viscoelasticity. Thixotropy is more pronounced at low shear rates with a
long time scale. The effect, however, seem to have a less important role in blood flow than other
non-Newtonian effects such as shear thinning \cite{Mandal2005}, and this could partly explain the
limited amount of studies dedicated to this property. The thixotropic behavior of blood is very
sensitive to the blood composition and hence it can demonstrate big variations between different
individuals and under different biological conditions \cite{Dintenfass1962}.

It should be remarked that time dependent effects in general, whether thixotropic or viscoelastic
in nature or of any other type, should be expected in the flow of blood due to a number of reasons.
One reason is the pulsatility of blood flow and the rapid change in the deformation conditions
during the systolic-diastolic cardiac cycle. Another reason is the rapid change in the shear
magnitude between one part of the system to another part, i.e. different shear rates between the
arteries, capillaries, porous tissue and venous part \cite{AlonsoPG1993}. A third reason is the
irregular shape, such as bends and converging-diverging formations, of the blood flow conduits
\cite{SochiPower2011, SochiNavier2013} which activates or accentuates time-dependent effects. A
fourth reason is the difference in the deformation rates between the ventricular systole and
diastole. The last reason may explain the indication of one study \cite{FisherR2009} that the
non-Newtonian effects are more important at diastole than systole since the shear rates during
diastole are expected to be lower than those at systole.

Another remark is that most of the reported non-Newtonian rheological parameters, as well as many
other physical properties of blood, are obtained from {\it in vitro} measurements and hence they
are subject to significant errors as an indicator to the {\it in vivo} values due to the difference
in ambient conditions as well as the experimental requirements and procedures, such as using
additives to preserve and fluidize the blood samples, that can introduce significant variations on
the blood properties. Moreover, the reported values could be highly dependent on the measurement
method \cite{LeeXNLS2011}. The differences between the individual subjects and their conditions
like dietary intake prior to measurement \cite{VlastosTR2003}, most of which are difficult to
control or quantify, should add to the uncertainties and fluctuations. Hence, most of these values
should be considered with caution especially when used for {\it in vivo} and patient-specific
modeling and investigation.

%XXXXXXXXXXXXXXXXXXXXXXXXXXXXXXXXXXXXXXXXXXXXXXXXXXXXXXXXXXXXXXXXX
\section{Modeling non-Newtonian Effects}\label{ModelingSec}

Blood is a complex non-Newtonian fluid and hence reliable modeling of blood flow in the circulation
system should take into account its non-Newtonian characteristics. Several non-Newtonian
rheological models have been used to describe the blood rheology; these models include
Carreau-Yasuda \cite{GijsenVJ1998, GijsenAVJ1999, BoxGRR2005, ChenLW2006, JonasovaV2008,
VimmrJ2008, KimVL2008, LukacovaZ2008, FisherR2009, SankarI2009, ZauskovaM2010, LiuT2011,
MollaP2012},
Casson \cite{Merrill1969, ShuklaPR1980, MorrisSB1987, MorrisRSGSB1989, LouY1993, DasJP2000,
BoydBG2007, GapinskaGEJK2007, KimVL2008, FanJZLCD2009, FisherR2009, LeeXNLS2011, MollaP2012,
HuangCS2013},
power law \cite{ShuklaPR1980, JohnstonJCK2006, RevellinRBB2009, FisherR2009, LiuT2011,
BodnarSP2011, MollaP2012},
Cross \cite{AbrahamBH2005, JonasovaV2008, VimmrJ2008, KimVL2008, MollaP2012},
Herschel-Bulkley \cite{ValenciaZGB2006, SankarI2009, VajraveluSDP2011, LeeXNLS2011, MisraM2012},
Oldroyd-B \cite{Pontrelli1998, PerktoldKLH1999, Pontrelli2000, BodnarSP2011},
Quemada \cite{DasEP1997, GapinskaGEJK2007, KimVL2008, MollaP2012},
Yeleswarapu \cite{YilmazG2008, LukacovaZ2008, ZauskovaM2010},
Bingham \cite{LouY1993},
%
%Jeffrey \cite{AkbarN2012},
%
Eyring-Powell \cite{ZuecoB2009}, and
Ree-Eyring \cite{GapinskaGEJK2007}.
The constitutive equations of these rheological models are given in Table \ref{BloodModelsTable}.
Other less known fluid models \cite{IshikawaGOY1998} have also been used to describe the rheology
of blood. A quick inspection of the blood literature reveals that the most popular models in
non-Newtonian hemorheologic and hemodynamic modeling are Carreau-Yasuda and Casson.

%%%%%%%%%%%%%%%%%%%%%%%%%%%%%%%%%%%%%%%%%%%%%%%%%%%%%%%%%%%%%%%%%%%%%%%%%%%%%%%%%%%%%%%%%

\begin{table} [!t]
\caption{The non-Newtonian fluid models that are commonly used to describe blood rheology. The
meanings of symbols are given in Nomenclature \S\ \ref{Nomenclature}. The symbols that define fluid
characteristic properties, such as $\lambda$, are generically used and hence the same symbol may
represent different physical attributes in different models. Some of these models may have more
than one form; the one used in this table is the widespread of its variants. The last column
represents the frequently obtained non-Newtonian properties from these models in the context of
blood modeling although other properties may also be derived and employed in modeling other
materials.} \label{BloodModelsTable} \vspace{0.2cm} \centering
\begin{tabular}{|l|l|l|}
\hline
\textbf{Model} & \textbf{Equation} & \textbf{Non-Newtonian Properties}\tabularnewline
\hline
Carreau-Yasuda & $\mu=\mu_{\infty}+\frac{\mu_{0}-\mu_{\infty}}{\left[1+\left(\lambda\dot{\gamma}\right)^{a}\right]^{\frac{1-n}{a}}}$ & shear thinning\tabularnewline
\hline
Casson & $\tau^{1/2}=\left(k\dot{\gamma}\right)^{1/2}+\tau_{o}^{1/2}$ & yield stress\tabularnewline
\hline
Power law & $\tau=k\dot{\gamma}^{n}$ & shear thinning\tabularnewline
\hline
Cross & $\mu=\mu_{\infty}+\frac{\mu_{0}-\mu_{\infty}}{1+\lambda\dot{\gamma}^{m}}$ & shear thinning\tabularnewline
\hline
Herschel-Bulkley & $\tau=k\dot{\gamma}^{n}+\tau_{o}$ & shear thinning, yield stress\tabularnewline
\hline
Oldroyd-B & ${\sTen}+\rxTim{\ucd\sTen}=\lVis\left({\rsTen}+\rdTim{\ucd\rsTen}\right)$ & viscoelasticity\tabularnewline
\hline
Quemada & $\mu=\mu_{p}\left(1-\frac{k_{0}+k_{\infty}\sqrt{\dot{\gamma}/\dot{\gamma_{c}}}}{2\left(1+\sqrt{\dot{\gamma}/\dot{\gamma_{c}}}\right)}\phi\right)^{-2}$ & shear thinning\tabularnewline
\hline
Yeleswarapu & $\mu=\mu_{\infty}+\left(\mu_{0}-\mu_{\infty}\right)\frac{1+\ln\left(1+\lambda\dot{\gamma}\right)}{1+\lambda\dot{\gamma}}$ & shear thinning\tabularnewline
\hline
Bingham & $\tau=k\dot{\gamma}+\tau_{o}$ & yield stress\tabularnewline
\hline
Eyring-Powell & $\mu=\mu_{\infty}+\frac{\left(\mu_{0}-\mu_{\infty}\right)\mathrm{sinh^{-1}\left(\lambda\dot{\gamma}\right)}}{\lambda\dot{\gamma}}$ & shear thinning\tabularnewline
\hline
Ree-Eyring & $\tau=\mathrm{\tau_{c}\, sinh}^{-1}\left(\frac{\mu_{0}\dot{\gamma}}{\tau_{c}}\right)$ & shear thinning\tabularnewline
\hline
\end{tabular}
\end{table}

%%%%%%%%%%%%%%%%%%%%%%%%%%%%%%%%%%%%%%%%%%%%%%%%%%%%%%%%%%%%%%%%%%%%%%%%%%%%%%%%%%%%%%%%%

Blood is also modeled as a Newtonian fluid \cite{JohnstonJCK2006, FanJZLCD2009, LiuT2011,
BodnarSP2011} which is a good approximation in many circumstances such as the flow in large vessels
at medium and high shear rates under non-pathological conditions \cite{BoydBG2007}. As there is no
sudden transition from non-Newtonian to Newtonian flow as a function of shear rate, there is no
sharply-defined critical limit for such a transition \cite{YilmazG2008} and hence this remains a
matter of choice which depends on a number of objective and subjective factors. However, there
seems to be a general consensus that the shear rate range for which non-Newtonian effects are
considered significant is $\lesssim100$~s$^{-1}$ \cite{ReplogleMM1967, LouY1993, BaskurtM2003,
JohnstonJCK2006, MollaP2012}; above this limit the blood is generally treated as a Newtonian
liquid.

No single model, Newtonian or non-Newtonian, can capture all the features of the blood complexities
\cite{YilmazG2008} and hence different models are used to represent different characteristics of
the blood rheology. These models, whether Newtonian or non-Newtonian, obviously have significant
differences and hence they can produce very different results \cite{AbrahamBH2005, LukacovaZ2008,
ZauskovaM2010}. The results also differ significantly between Newtonian and non-Newtonian models in
most cases \cite{VimmrJ2008, MollaP2012}. The non-Newtonian models vary in their complexity and
ability to capture different physical phenomena.

Diverse methods have been used in modeling and simulating non-Newtonian effects in blood rheology;
these include analytical \cite{MorrisSB1987, VajraveluSDP2011, MisraM2012}, stochastic
\cite{OuaredC2005, BoydBG2007, HuangCS2013, FuLS2013a, FuLS2013b}, and numerical mesh methods; such
as finite element \cite{BoxGRR2005, AbrahamBH2005, CanicHRTGM2006, ChenLW2006, ChapelleGMC2010},
finite difference \cite{LouY1993, IshikawaGOY1998, Pontrelli1998, Pontrelli2000, ZauskovaM2010,
MandalML2012}, finite volume \cite{JonasovaV2008, VimmrJ2008, FisherR2009, FanJZLCD2009,
ZauskovaM2010, BodnarSP2011, MollaP2012}, and spectral collocation methods \cite{Pontrelli1998,
Pontrelli2000}.

As indicated previously, most, if not all, non-Newtonian characteristics arise from the blood
microstructure and particularly the concentration, distribution and mechanical properties of the
red blood cells. For example, the viscoelastic properties of blood originate from the mechanical
properties of the suspended cells and their capability of elastic deformation and structural
formation, while the thixotropic properties arise from steady disaggregation of blood cells over
prolonged shearing time. Interestingly, the majority of the rheological models used to describe the
blood rheology are bulk phenomenological models of empirical nature with little consideration, if
any, to its highly influential micro-structure. Hence, more structurally-based models, such as
those based on molecular dynamics, are required to improve the description and modeling of the
blood rheological behavior.

%XXXXXXXXXXXXXXXXXXXXXXXXXXXXXXXXXXXXXXXXXXXXXXXXXXXXXXXXXXXXXXXXX
\section{Non-Newtonian Effects in Circulation Subsystems}

Because the impact of the non-Newtonian effects is highly dependent on the shape and size of the
flow conduits, different non-Newtonian rheological behavior, and hence different flow modeling
approaches, should apply to the different parts of the circulatory system. Different approaches are
also required because of the difference in the nature of the blood transportation processes in
these parts, such as large scale bulk flow in the large vessels as opposite to perfusion or
diffusion in the porous tissue.

\begin{figure} %[!h]
\centering %
\subfigure[Large Vessels]%
{\begin{minipage}[b]{0.5\textwidth} \centering \includegraphics[width=1.7in] {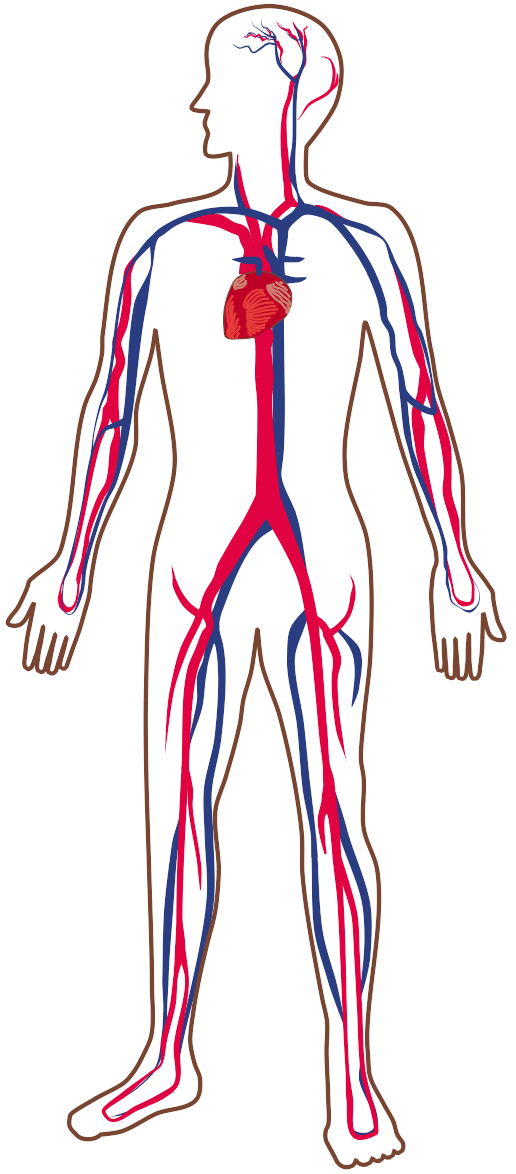}
\end{minipage}} \vspace{0.2cm}

%\Hs %
%XXXXXXXXXXXX
%
\centering %

\subfigure[Small Vessels]%
{\begin{minipage}[b]{0.5\textwidth} \centering \includegraphics[width=1.9in] {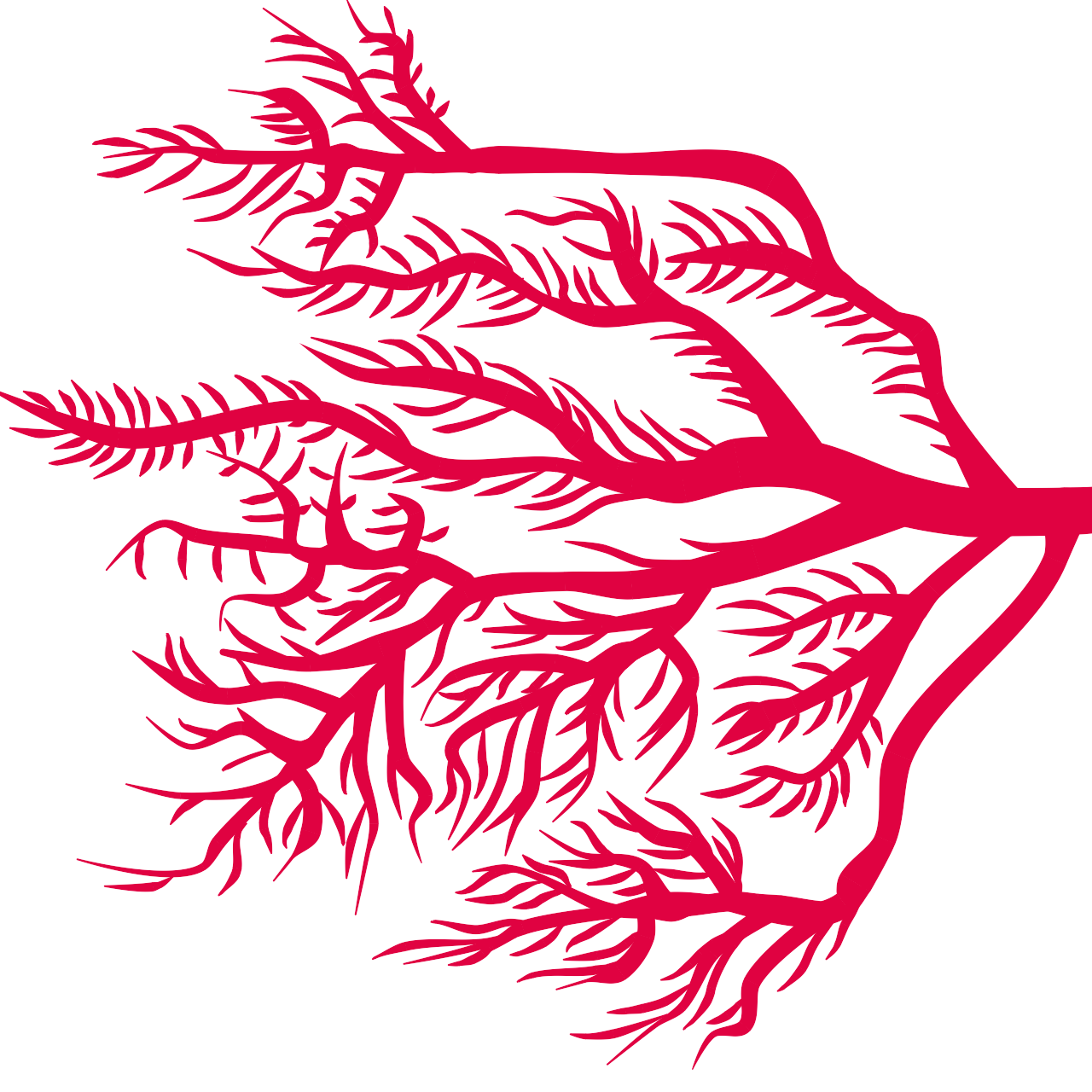}
\end{minipage}} \vspace{0.2cm}

%XXXXXXXXXXXX
%
\centering %
\subfigure[Porous Tissue]%
{\begin{minipage}[b]{0.5\textwidth} \centering \includegraphics[width=2.2in] {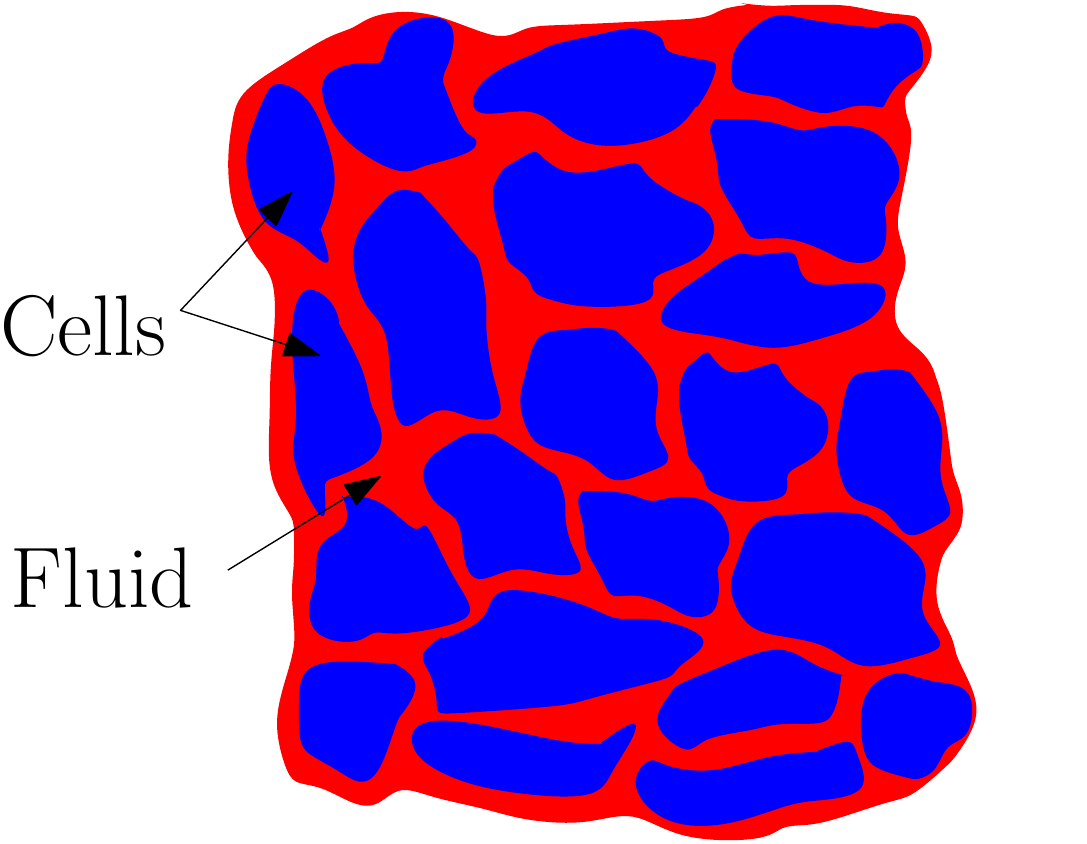}
\end{minipage}}
\caption{The three types of circulatory subsystems: (a) large vessels, (b) small vessels, and (c)
porous tissue. The three sub-figures are obviously drawn on different scales. \label{SubsystemFig}}
\end{figure}

We can identify three types of circulatory subsystems in which non-Newtonian effects should be
analyzed and modeled differently: large blood vessels which mainly apply to arteries and veins,
small blood vessels which broadly include capillaries and possibly arterioles, and porous tissue
such as the myocardium and muscles in general. These three subsystems are graphically illustrated
in Figure \ref{SubsystemFig}. The distinction between large and small vessels is not clear cut as
it depends on the nature of the flow phenomenon under consideration and the associated
circumstances. However, the distinctive feature that should be used as a criterion to differentiate
between these two categories of blood vessels in this context is the validity of the continuum
approximation as applied to the blood where in the large vessels this approximation is strictly
held true while in the small vessels it approaches its limits as some non-continuum phenomena start
to appear.

In the following subsections we outline general strategies for modeling non-Newtonian effects in
the circulation subsystems.

%%%%%%%%%%%%%%%%%%%%%%%%%%%%%%%%%%%%%%%%%%%%%%%%%%%%%%%%%%%%%%%%%%%%%%%%%%%%%%%%%%%%%%%%%

%XXXXXXXXXXXXXXXXXXXXXXXXXXXXXXXXXXXXXXXXXXXXXXXXXXXXXXXXXXXXXXXXX
\subsection{Large Vessels}

In large vessels, which include large cavities such as the ventricles and atria inside the
myocardium as well as the large arteries and veins, the blood essentially behaves as a Newtonian
fluid. One reason is that the blood in such large lumens and cavities is normally exposed to a
relatively high shear rates and hence the non-Newtonian effects which are basically induced at low
shear rates die out \cite{PerktoldKLH1999}. Also at this large scale the blood appears as a
homogeneous continuum medium with diminishing effect of blood cell aggregation \cite{GijsenVJ1998}.
The interaction between the blood cells with their pronounced elastic properties is also minimal at
this scale. However, in some pathological situations, non-Newtonian effects are important even in
the big cavities and large vessels and therefore they should be considered in the flow model. This
may also be true in some non-pathological situations in which the non-Newtonian effects can be
critical to the observed phenomena.

It should be remarked that the non-Newtonian effects in the venous part of the circulatory system
should be more important than the arterial part due to the lower deformation rates in the former
than the latter as this seems to be an accepted fact in the hemodynamic studies. However, we did
not find a proper discussion about this issue in the available literature. The difference in the
blood composition in these two parts; due for example to the difference in concentration of
substances like nutrients, oxygen and metabolic wastes; should also affect the non-Newtonian
rheology in these two subsystems and introduce more complications in modeling blood flow especially
in large vessels. Like the previous issue, we did not find an explicit discussion to this issue in
the available literature. Another remark is that certain parts of the large vessels network can
contain spots of low shear rates such as bends and bifurcation junctions \cite{LouY1993,
FisherR2009} and hence non-Newtonian effects in large vessels could be significant in some cases
where these spots play an exceptionally important role in the blood flow due to a diseased case for
instance.

Several mathematical and computational models have been used to describe the flow of blood in large
individual vessels. These models include the elastic one-dimensional Navier-Stokes and the rigid
Hagen-\pois\ for Newtonian fluids, as well as many other non-Newtonian rheological models such as
Cross and Carreau-Yasuda, as discussed previously is section~\ref{ModelingSec}. The characteristics
of blood flow in large single vessels are obtained from these mathematical models either
analytically or numerically, e.g. through the use of finite element or finite difference
techniques. Most of the employed non-Newtonian fluid models are generalized Newtonian models and
hence they do not account for history-dependent elastic or thixotropic effects. Also, the
analytical non-Newtonian models generally apply to rigid tubes only although there are some
attempts in this context to extend \pois\ flow to elastic vessels with non-Newtonian rheology
\cite{VajraveluSDP2011}. Numerical methods may also be used to extend the non-Newtonian models to
elastic vessels.

With regards to the flow in vascular networks of large vessels, the main models used to describe
and simulate blood flow are the one-dimensional Navier-Stokes finite element model for elastic
networks \cite{SochiTechnical1D2013, SochiElastic2013} and the Hagen-\pois\ model for rigid
networks \cite{SochiPois1DComp2013}. Both of these models are Newtonian although the second may be
extended to \pois-like non-Newtonian flow through the inclusion of time-independent non-Newtonian
effects using a vessel-dependent non-Newtonian effective viscosity which is computed and updated
iteratively to reach a consistent flow solution over the whole network \cite{Sochithesis2007,
SochiB2008, SochiComp2010, SochiFeature2010}. Based on a non-thorough inspection of the available
literature, there seems to be no extension to the traditional one-dimensional Navier-Stokes
distensible network model \cite{FormaggiaLQ2003, SherwinFPP2003, SochiTechnical1D2013} to
incorporate non-Newtonian effects. In fact we did not find serious attempts in the available
literature to extend the Navier-Stokes equation in general (whether one-dimensional or
multi-dimensional, for rigid or elastic conduits) to account for non-Newtonian effects, although
there are attempts to incorporate non-Newtonian effects numerically into Navier-Stokes flow models
\cite{XuCJ1997, ProkopK2010}. The Navier-Stokes equation with its nonlinearity is sufficiently
complex to be solved for Newtonian flow in most cases let alone with the added complexities and
nonlinearities introduced by the non-Newtonian rheology.

With regards to the one-dimensional Navier-Stokes distensible model for single vessels and
networks, we propose two possible general approaches for extending this model to account for
non-Newtonian effects. One approach is to accommodate these effects in the fluid viscosity as
parameterized by the viscosity friction coefficient $\kappa$. A second possible approach is to
incorporate these effects in the flow profile as described by the momentum flux correction factor
$\alpha$ in the one-dimensional model \cite{SochiTechnical1D2013, SochiElastic2013}. For the
network, the second approach is based on defining a vessel-dependent $\alpha$ field over the whole
network. A similar viscosity field may also be required for the first approach as well.

%XXXXXXXXXXXXXXXXXXXXXXXXXXXXXXXXXXXXXXXXXXXXXXXXXXXXXXXXXXXXXXXXX
\subsection{Small Vessels}

Non-Newtonian effects are generally more pronounced in small flow ducts, such as capillaries, than
in large ducts like arteries due to several reasons such as the deterioration of the continuum
assumption at small scales especially for complex dispersed systems like blood. In such vessels the
continuum approximation reaches its limit and the effect of blood cell aggregation with their
interaction with the vessel wall becomes pronounced. This activates the non-Newtonian rheological
flow modes such as the induction of elastic effects which are associated with the elastic
properties of the red blood cells and their structural formation. Also, the non-Newtonian effects
of blood are more prominent at low shear rates which are the dominant flow regimes in the small
vessels \cite{KimVL2008}. Hence non-Newtonian rheological effects should be considered in modeling,
simulating and analyzing the flow of blood in small vessels.

%XXXXXXXXXXXXXXXXXXXXXXXXXXXXXXXXXXXXXXXXXXXXXXXXXXXXXXXXXXXXXXXXX
\subsection{Porous Tissue}

The commonly used approach in modeling blood perfusion in living tissue is to treat the tissue as a
spongy porous medium and employ Darcy law \cite{HuygheACR1992, KhaledV2003, CoussyBook2004,
ChapelleGMC2010, CoussyBook2010, SochiArticle2010} which correlates the volumetric flow rate to the
pressure gradient. There are several limitations in the Darcy flow model in general and in its
application to the blood flow through biological tissue in particular, and hence remedies have been
proposed and used to improve the model. Since this law is originally developed for the flow through
rigid porous media, modified versions are normally used to account for elasticity as required for
modeling biological tissue \cite{VankanHJHHS1997, ChapelleGMC2010}. Other limitations include
neglecting edge effects and the restriction imposed by the laminar low velocity assumption on which
the Darcy flow is based. The former may be overcome by employing the boundary term in the Brinkman
equation while the latter can be eliminated through the use of the Forchheimer model which
incorporates a high-velocity inertial term \cite{KhaledV2003}.

Since Darcy law in its original formulation is based on the Newtonian flow assumptions,
non-Newtonian rheology is generally ignored in the modeling of blood perfusion through porous
tissue. There have been several extensions and modifications to the Darcy law to include
non-Newtonian effects in the flow of fluids in general, and polymers in particular, through rigid
non-biological porous media. These attempts include, for example, viscoelastic models
\cite{SadowskiB1965, GogartyLF1972, HaroRW1996, Garrouch1999}, Herschel-Bulkley
\cite{AlfarissP1984}, power law \cite{FadiliTP2002, AlnimrA2004}, Blake-Kozeny-Carman
\cite{KozickiT1988}, as well as other non-Newtonian prototypes \cite{KondicMS1996, KondicSM1998}.
Pore scale network modeling has also been used to accommodate various non-Newtonian effects; such
as shear thinning, yield stress and viscoelasticity; in the flow of polymers through rigid porous
media \cite{SorbieCJ1989, Lopezthesis2004, Balhoffthesis2005, PerrinTSC2006, Sochithesis2007,
SochiB2008, SochiVE2009, SochiYield2010, SochiComp2010}. Similarly, other computational techniques,
such as stochastic lattice Boltzmann \cite{BoekCC2003, SullivanGJ2006}, have been tried to simulate
and investigate the non-Newtonian effects of the flow through rigid porous media in non-biological
studies. However, it seems there is hardly any work on modeling the non-Newtonian effects in the
blood flow through living tissues by incorporating these effects into the distensible Darcy flow
model.

To conclude, non-Newtonian effects in the blood perfusion through porous tissue are not negligible
in general due to the existence of fluid shearing and extensional forces which activate
non-Newtonian rheology. As the deformation rates in this type of transportation is normally low,
and considering the small size and tortuous converging-diverging shape of the porous space inside
which the blood perfuses, non-Newtonian rheological effects are expected to be significant.
Non-Newtonian rheological effects associated with other fluid transport phenomena, such as
diffusion, could be negligible in such porous space due to the absence of shearing and extensional
forces as a result of lack of large scale fluid bulk movement in these micro- and nano-scale
phenomena. However, the causes underlying the non-Newtonian rheology should affect these transport
phenomena as well, although more serious investigations are required to reach any definite
conclusion about these issues. The existing literature is, unfortunately, limited in this scope
\cite{KumarU1980, CummingsWEF1991}.

%XXXXXXXXXXXXXXXXXXXXXXXXXXXXXXXXXXXXXXXXXXXXXXXXXXXXXXXXXXXXXXXXX
\section{Conclusions} \label{Conclusions}

Blood is essentially a non-Newtonian suspension. Its complex rheological non-Newtonian behavior is
largely influenced by its microstructure which, through the essentially viscous water-based
Newtonian plasma combined with the effect of aggregation, deformation and orientation of the
suspended blood cells with their distinguished elastic and three-dimensional structural formation
properties, can show diverse non-Newtonian effects at various shear rate regimes and through
different flow conduit structures although these effects are more pronounced at certain flow
regimes and in particular structures.

Blood rheological properties, and its mechanical characteristics in general, are affected by
several factors such as the type and magnitude of deformation rate, hematocrit level and protein
concentration. Because blood is a suspension, its properties can be strongly influenced by the
shape and size of its flow conduits. The non-Newtonian effects of blood can be accentuated by
certain pathological conditions such as hypertension and myocardial infarction.

Apart from some extreme diseased states where the non-Newtonian effects play an exceptionally
important role in the fluid transport phenomena, the non-Newtonian effects are generally mild in
the bulk flow of blood in large vessels. More important influence of non-Newtonian rheology occurs
in the flow of blood through small vessels and in the blood perfusion through porous tissue. Other
transport phenomena, like diffusion, which do not involve bulk flow associated with deformation
forces of shearing or extensional type that activate non-Newtonian rheology, should not be affected
directly by the non-Newtonian characteristics although the physical causes at the root of the
non-Newtonian rheology should have an impact on these processes. More fundamental studies are
required to reach specific conclusions about these issues.

Non-Newtonian rheology, such as viscoelasticity, of the blood vessel walls and the spongy porous
tissue should also be considered as a contributor to the overall non-Newtonian behavior of blood
circulation as these effects are both non-Newtonian and circulatory in nature like the ones
demonstrated by the blood itself. The effect of fluid-structure interaction should also be included
in analyzing, modeling and simulating of non-Newtonian effects in the blood circulation as it plays
an important hemodynamic and hemorheologic role.

%\clearpage
\phantomsection \addcontentsline{toc}{section}{References} %
\bibliographystyle{unsrt}
%\bibliography{Bibl}

\clearpage
%XXXXXXXXXXXXXXXXXXXXXXXXXXXXXXXXXXXXXXXXXXXXXXXXXXXXXXXXXXXXXXXXX
\section{Nomenclature}\label{Nomenclature}

\begin{supertabular}{ll}
$\sR$                   & shear rate \\
$\sR_c$                 & characteristic shear rate \\
$\rsTen$                & rate of strain tensor \\
$\lambda$               & characteristic time constant \\
$\rxTim$                & relaxation time \\
$\rdTim$                & retardation time \\
$\Vis$                  & fluid viscosity \\
$\Vis_p$                & plasma viscosity \\
$\lVis$                 & zero-shear-rate viscosity \\
$\Vis_\infty$           & infinite-shear-rate viscosity \\
$\sS$                   & shear stress \\
$\sTen$                 & stress tensor \\
$\sS_c$                 & characteristic shear stress \\
$\ysS$                  & yield-stress \\
$\phi$                  & volume concentration \\
\\
$a$                     & Carreau-Yasuda index \\
$k$                     & consistency coefficient \\
$k_0$                   & maximum volume fraction for zero shear rate \\
$k_\infty$              & maximum volume fraction for infinite shear rate \\
$m$                     & Cross model index \\
$n$                     & power law index \\
$\ucd \cdot$            & upper convected time derivative \\

\end{supertabular}

\end{document}